\documentclass[a4paper,11pt]{article}
\usepackage{pos}
\title{POEMMA (Probe of Extreme Multi-Messenger Astrophysics) Roadmap Update}
 \ShortTitle{POEMMA Update}

\author*[a]{Angela V. Olinto}

\affiliation[a]{The University of Chicago, Department of Astronomy \& Astrophysics,\\
  5640 South Ellis Avenue, Chicago, IL, USA}

\onbehalf{for the POEMMA and JEM-EUSO collaborations} 

\emailAdd{aolinto@uchicago.edu}

\abstract{
The Probe Of Extreme Multi-Messenger Astrophysics (POEMMA) was designed as a NASA Astrophysics probe-class mission to identify the sources of ultrahigh energy cosmic rays (UHECRs) and observe cosmic neutrinos from extremely energetic transient sources. POEMMA consists of two identical spacecraft flying in a loose formation at 525 km altitude oriented to view a common atmospheric volume and to provide full-sky coverage for both types of messengers. Each spacecraft hosts a wide field of view Schmidt telescope with a hybrid focal plane optimized to observe both the UV fluorescence signal from extensive air showers (EASs) and the optical Cherenkov signals from EASs. When in stereo close to nadir mode, POEMMA can measure the spectrum, composition, and full-sky distribution of the UHECRs above 20 EeV and be sensitive to UHE neutrinos. When pointing just below the Earth’s limb, POEMMA will be sensitive to cosmic tau neutrinos above 20 PeV by observing the Cherenkov radiation of EASs produced by upward-moving tau decays, induced from tau neutrino interactions in the Earth. POEMMA is designed to quickly re-orient to follow a Target-of-Opportunity (ToO) neutrino transient from astrophysical sources with exceptional sensitivity to neutrinos from both short-duration transients, such as short-gamma-ray bursts (sGRBs), and long-duration sources, such as binary neutron star (BNS) mergers. Here we review the POEMMA mission and discuss the recent progress towards its technical readiness  provided by the Mini-EUSO and EUSO-SPB2 missions and the forthcoming Terzina and POEMMA-Balloon-Radio missions.
}

\ConferenceLogo{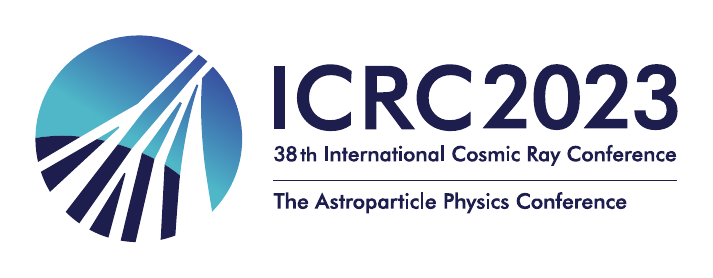}

\FullConference{%
38th International Cosmic Ray Conference (ICRC2023)\\
  26 July - 3 August, 2023\\
  Nagoya, Japan}

\begin{document}
\maketitle

\section{POEMMA Science Goals}
\begin{figure}
\begin{center}
\includegraphics [width=1\textwidth]{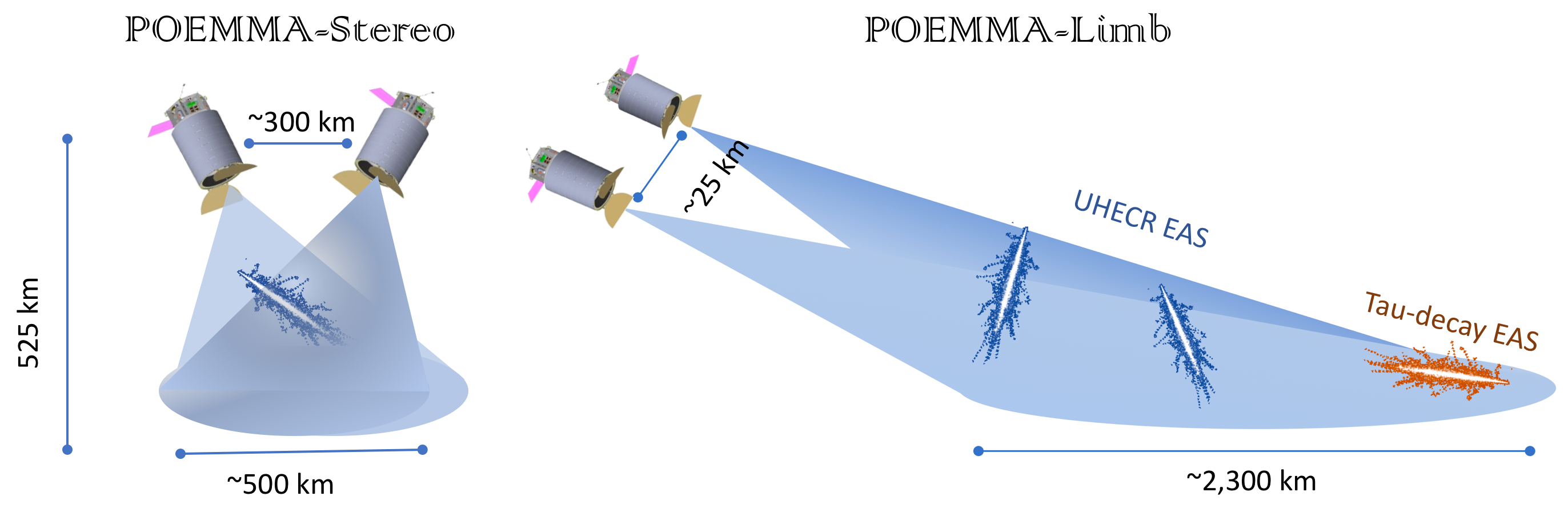}
\caption{POEMMA observing modes. {\it Left:} POEMMA-Stereo mode to observe fluorescence from UHE cosmic rays and neutrinos in stereo. (Telescope separation $\sim$300 km and pointing close to nadir for the most precise measurements at 10s of EeV.) {\it Right:} POEMMA-Limb mode to observe Cherenkov from cosmic neutrinos just below the limb of the Earth and fluorescence from UHECRs throughout the atmospheric volume. (Telescope separation $\sim$25 km and pointing towards rising or setting ToO sources.)}
\label{fig1}
\end{center}
\end{figure}

As described in ~\cite{POEMMA_JCAP_2021}, the main scientific goals of  POEMMA  are to discover the elusive sources of cosmic rays with energies above 10$^{18}$ eV ($\equiv$ 1 EeV) and to observe cosmic neutrinos with energies above 20 PeV from multi-messenger transients. POEMMA exploits the tremendous gains in both ultrahigh energy cosmic ray (UHECR) and cosmic neutrino exposures offered by space-based measurements, including the {\it full-sky coverage} of the celestial sphere.  For cosmic rays with energies $E \gtrsim 20~{\rm EeV}$, POEMMA can measure the UHECR spectrum,  composition, and source location with high statistics at the highest energies. For multi-messenger transients, POEMMA can follow-up targets of opportunity (ToO) to detect cosmic neutrinos with energies $E_{\nu} \gtrsim$ 20 PeV. 
POEMMA also has sensitivity to neutrinos with energies above 20~EeV through fluorescence observations of neutrino induced EASs. Supplementary science capabilities of POEMMA include probes of physics beyond the Standard Model of particle physics, the measurement of $pp$ cross-section at $\sim$ 0.3 PeV center-of-mass energy, the study of atmospheric transient luminous events (TLEs), and the search for meteors and nuclearites. 

\begin{figure}
\begin{center}
\includegraphics [width=1\textwidth]{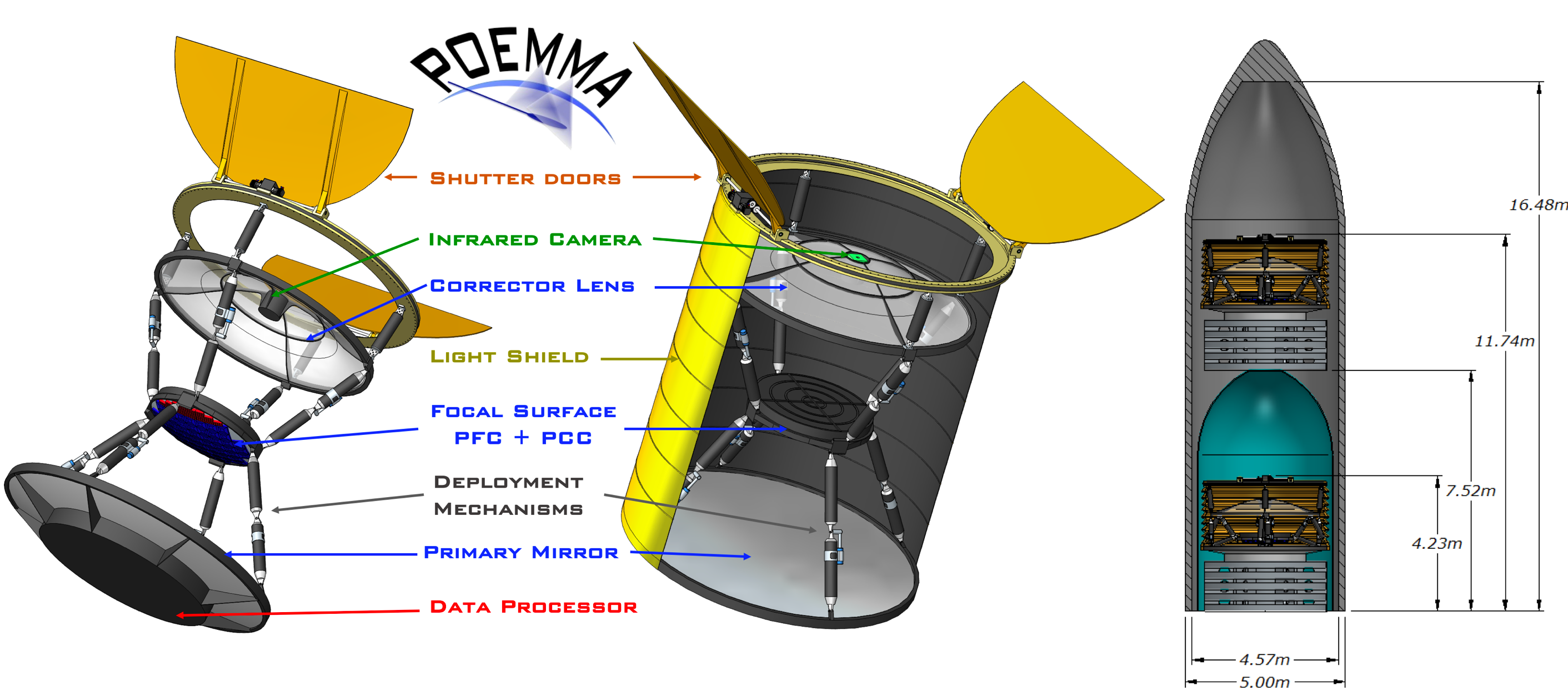}
\caption{{\it Left:} Concept of the POEMMA photometer with major components identified (PFC stands for POEMMA Fluorescence Camera and PCC stands for POEMMA Cherenkov Camera). {\it Right:} Both POEMMA photometers accommodated on Atlas V for launch. From~\cite{POEMMA_JCAP_2021}.}
\label{fig2}
\end{center}
\end{figure}

POEMMA can achieve this significant increase in sensitivity by operating two observatories (described in Fig.\ref{fig2} and Table I) with very wide field of view (FoV) in different orientation modes: a stereo fluorescence configuration pointing close to the nadir for more precise UHECR observations and a tilted, Earth-limb viewing configuration to follow ToO neutrino searches (see Fig.\ref{fig1}). In limb observing mode, POEMMA can simultaneously search for neutrinos  with Cherenkov observations, while observing UHECRs with fluorescence, thanks to the POEMMA hybrid focal surface design.  

POEMMA's fluorescence observations can yield one order of magnitude increase in yearly UHECR exposure compared to ground observatory arrays and two orders of magnitude compared to ground fluorescence telescopes. In the limb-viewing mode, POEMMA searches for optical Cherenkov signals of upward-moving EASs generated by $\tau$-lepton decays produced by $\nu_\tau$ interactions in the Earth with a terrestrial neutrino target of  $\sim 10^{10}$ gigatons. In the limb-viewing mode, an even more extensive volume can be monitored for UHECR fluorescence observations.

\begin{table}
\centering
\caption{POEMMA Specifications:}
\label{tab-1}  \
\begin{tabular}{lllllllll}
\hline
\hline
Photometer & Components &  &$\ \ $& Spacecraft  & \\ \hline
Optics &  Schmidt & 45$^\circ$ full FoV && Slew rate & 90$^\circ$  in 8 min \\
 & Primary Mirror & 4 m diam. && Pointing Res. & 0.1$^\circ$ \\
 & Corrector Lens & 3.3 m diam. && Pointing Know. & 0.01$^\circ$ \\
 & Focal Surface & 1.6 m diam. && Clock synch. & 10 ns \\  
 & Pixel Size & $3 \times 3$ mm$^2$  && Data Storage & 7 days \\ 
& Pixel FoV & 0.084$^\circ$ && Communication & S-band \\ 
PFC & MAPMT (1$\mu$s)& 126,720 pixels  && Wet Mass & 3,450 kg \\
PCC & SiPM (20 ns)& 15,360 pixels  && Power (w/cont)& 550 W \\ \hline
Photometer & (One)&  &&Mission  &(2 Observatories) \\ \hline
 & Mass & 1,550 kg  && Lifetime & 3 year  (5 year goal)\\
 & Power (w/cont) & 700 W   && Orbit & 525 km, 28.5$^\circ$ Inc \\
 & Data & $<$ 1 GB/day && Orbit Period & 95 min \\
& &  && Observatory Sep. & $\sim$25 - 1000 km \\\hline
\hline
\end{tabular}
 \
\center{Each Observatory = Photometer + Spacecraft; POEMMA Mission = 2 Observatories}
\end{table}

\medskip
\noindent{\bf UHECR Science:} 

\begin{figure}
\begin{center}
\includegraphics [width=1\textwidth]{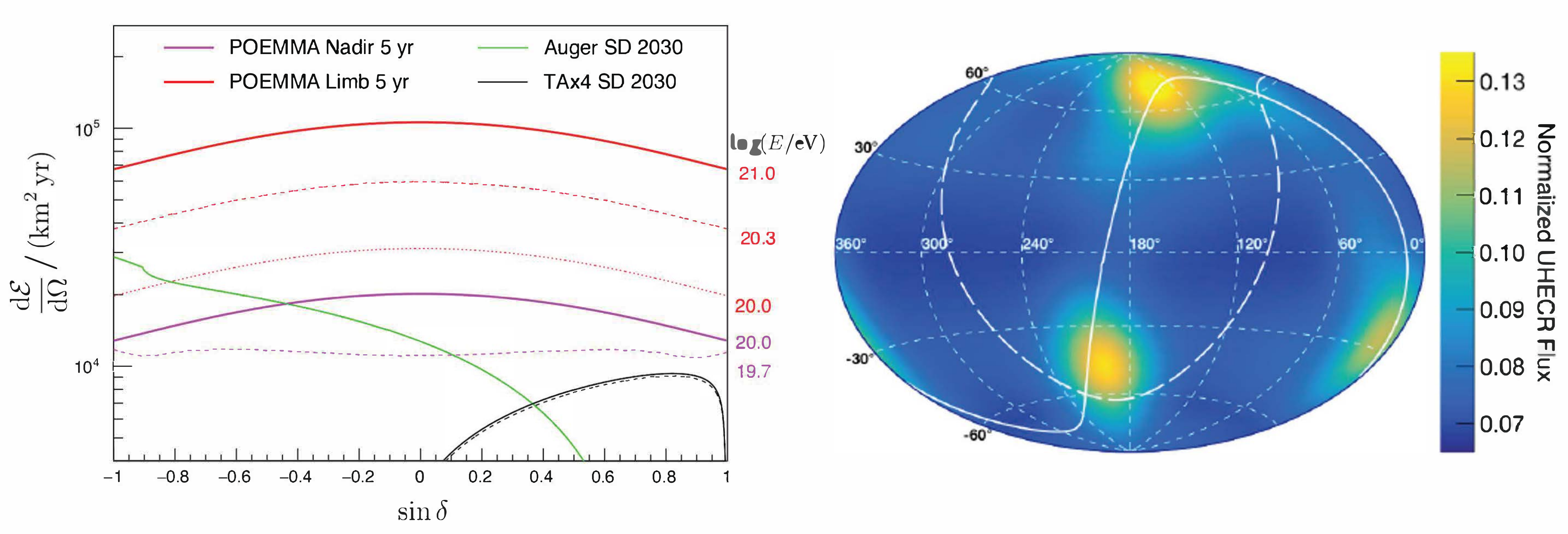}
\caption{{\it Left:}  Differential exposure vs declination for POEMMA 5-yr in nadir  (purple) at 10$^{19.7}$ eV (dotted) and 10$^{20}$ eV (solid); and for limb (red) at 10$^{20}$ eV (dotted), 10$^{20.3}$ eV (dashed), and 10$^{21}$ eV (solid).  Exposures for Auger (green) and TAx4  (black) surface detectors (SD)  until 2030.
 {\it Right:}  Sky map of the normalized UHECR flux in equatorial coordinates for starburst galaxies with 11\% anisotropy fraction and angular spread of $15^{\circ}$.  Adapted from~\cite{POEMMA_JCAP_2021}.}
\label{figUHECR}
\end{center}
\end{figure}

As summarized in ~\cite{UHECR_Snowmass}, the nature of UHECR sources and their acceleration mechanism(s) remain a mystery.  Proposed sources span a large range of astrophysical objects including extremely fast-spinning young pulsars, active galactic nuclei (AGN), starburst galaxies (SBGs),  gamma-ray bursts (GRBs), and galaxy clusters~\cite{UHECR_OpenQ}. 
The powerful exposure provided by POEMMA is designed to help determine the sources of UHECRs  through the combined detailed observations of the sky distribution, the spectrum, and the composition at the highest energies (above 100 EeV). Fig.~\ref{figUHECR} shows the POEMMA exposure compared to leading ground observatories on the left and a sky map in equatorial coordinates for starburst galaxies on the right.
 POEMMA can explore the differences in source models for the UHECRs by measuring the spectrum above the current reach in energy and the UHECR composition at 100s of EeV where models differ in predictions further illuminating the origin of UHECRs~\cite{POEMMA_JCAP_2021}.

\medskip
\noindent{\bf Cosmic Neutrino Science:}

Very high-energy cosmic neutrinos are emitted in a number of models of astrophysical transient events (gamma-ray bursts, blazars, binary neutron star coalescence, etc.). Astrophysical sources generally produce electron and muon neutrinos, which after astronomical propagation distances,  arrive on Earth with approximately equal numbers of the three flavors: electron, muon, and tau neutrinos. POEMMA detects primarily tau-neutrinos through the tau-decay generated EASs corresponding to about a  third of the generated neutrino flux. POEMMA has the unique capability of detecting neutrinos above 20 PeV (1 PeV $\equiv$ 10$^{15}$ eV)  from ToO  transient events with a follow-up  time scale of about one orbit (95 min) over the entire dark sky~\cite{POEMMA_JCAP_2021}. POEMMA can follow up these events and reach a neutrino fluence around $E_{\nu}^2J_{\nu} \ge  0.1 \ {\rm GeV}/ {\rm cm}^{-2}$ depending on the location of the sources. 
POEMMA can observe the full sky after several months, given the orbit of the Earth around the Sun. 

\begin{figure}
\begin{center}
\includegraphics [width=1\textwidth]{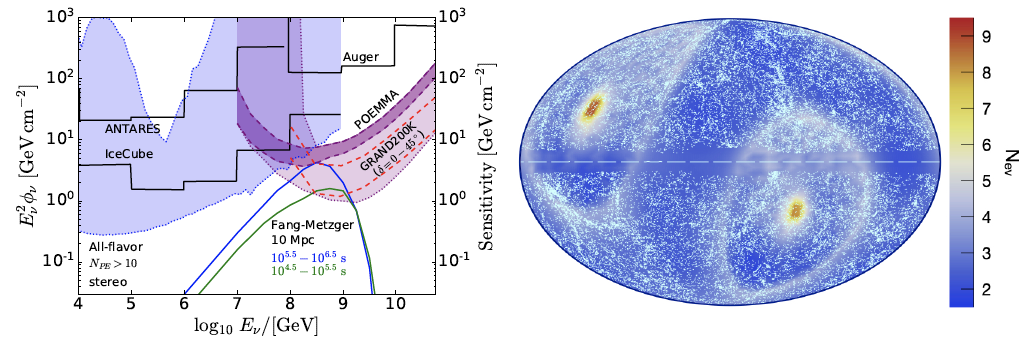}
\caption{{\it Left:} POEMMA ToO sensitivities to a long burst. The magenta bands (dark and light) show the range of possible source locations. 
Also shown are the IceCube, Antares, and Auger upper limits (solid histogram) from a neutrino search within a 14-day time window around the binary neutron star merger GW170817~\cite{ANTARES:2017bia}. The blue bands show the variation of IceCube sensitivity due to celestial source location and the red dashed curves represent the projected sensitivity of GRAND200k at zenith angles $90^\circ$ and $94^\circ$ \cite{Alvarez-Muniz:2018bhp},
and models from Fang \& Metzger~\cite{Fang:2017tla} of the all-flavor neutrino fluence produced $10^{5.5}-10^{6.5}$~s and $10^{4.5}-10^{5.5}$~s after a binary neutron star merger event occurring at a distance of 10~Mpc. 
{\it Right:} Sky plot of the expected number of neutrino events with POEMMA as a function of galactic coordinates for the Fang \& Metzger~\cite{Fang:2017tla} binary neutron star merger model, placing the source at $10$~Mpc. Figure adapted from~\cite{POEMMA_JCAP_2021} . }
\label{Neutr1}
\end{center}
\end{figure}

\section{POEMMA Instrument and Mission}

As shown in Fig.\ref{fig2} and Table I, the POEMMA observatory~\cite{POEMMA_JCAP_2021}  is comprised of two identical space-based platforms that detect extreme energy particles by recording the signals generated by EASs in the dark side of the Earth's atmosphere.  The central element of each POEMMA observatory  is a high sensitivity low resolution photometer that measures  two types of emission from these EASs: the faint isotropic emission due to the fluorescence of atmospheric nitrogen excited by air shower particles, and the brighter collimated Cherenkov emission from EASs directed at the POEMMA observatory. The photometers are designed for deployment after launch. A stowed configuration enables two identical satellites to be launched together on a single Atlas V rocket. Space qualified mechanisms extend each instrument after launch to their deployed position to begin observations.
The instrument architecture incorporates a large number of identical parallel sensor chains that meet the high standards of a Class B mission. 

The POEMMA {\bf photometer} is based on a Schmidt optical design with a large spherical primary mirror (4 m diameter), the aperture and a thin refractive aspheric aberration corrector lens (3.3 m diameter) at its center of curvature, and a convex spherical focal surface (1.61 m diameter). This particular system provides a large collection aperture (6.39 m$^2$) and a massive field-of-view (45$^\circ$ full FoV).   The diameter of the POEMMA primary mirror is set to fit the launch vehicle (Atlas V). The point-spread-function of the POEMMA optics is much less than a pixel size. POEMMA's imaging requirement is  $10^4$ away from the diffraction limit, implying optical tolerances closer to a microwave dish than an astronomical telescope. 
 
\begin{figure}
\begin{center}
\includegraphics [width=1\textwidth]{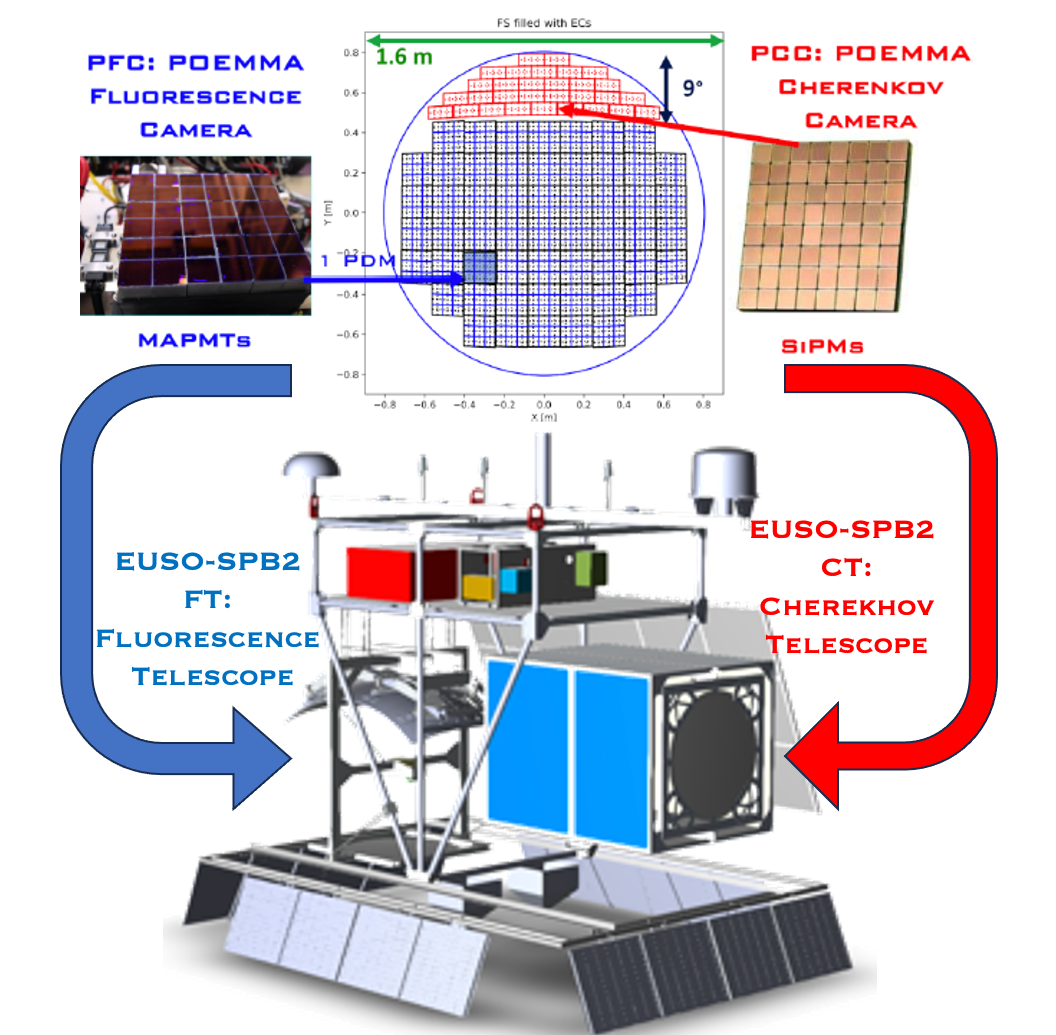}
\caption{{\it Top:}  POEMMA hybrid focal surface: POEMMA Fluorescence Camera (PFC) on the left and POEMMA Cherenkov Camera (PCC) on the right. The PFC records fluorescence from EAS in 1$\mu$s frames in the UV band using 1,980 MAPMTs with 3x3 mm$^2$ pixels to a total of 126,720 pixels. The PCC  uses SiPMs assembled in arrays of 8x8 pixels with a total of 15,360 pixels optimized to observe in the  optical band for Cherenkov emission with 20 nanosecond time integration. 
{\it Bottom:}   EUSO-SPB2 Telescopes~\cite{EUSO_SPB2_Overview}: 
The EUSO-SPB2 FT is a 1m diameter modified Schmidt telescope, with a 6,912 pixel camera of MAPMTs and integration time of 1$\mu$s. The FT field of view is 36$^\circ$ by 12$^\circ$ with a fixed nadir pointing direction.
The EUSO-SPB2 CT is also a 1m diameter modified Schmidt telescope with a bifocal alignment of the 4 mirror segments to reduce the background. The CT camera has a 512 pixel SiPM camera with an integration time of 10~ns. The FoV of the instrument is 6.4$^\circ$ in zenith and 12.8$^\circ$ in azimuth and can be pointed during the flight from horizontal to 10$^\circ$ below the Earth's limb to follow ToOs.
  }
\label{figComp}
\end{center}
\end{figure}

The {\bf focal surface} (FS) of the POEMMA photometer consists of the POEMMA Fluorescence Camera (PFC), optimized for the fluorescence signals, and the POEMMA Cherenkov Camera (PCC), optimized for Cherenkov signals (see Fig.~\ref{figComp}).
The PFC records the EAS videos in 1$\mu$s frames in the $300 \lesssim \lambda/{\rm nm} \lesssim 500$ wavelength band using multi-anode photomultiplier tubes (MAPMTs).  Each MAPMT has 64  (3x3 mm$^2$) pixels in an 8 x 8 array.  The PFC is composed of 1,980 MAPMTs containing a total of 126,720 pixels. The stereo videos of the EAS determines the energy, direction, and composition of the UHECR.
The PCC uses solid-state silicon photomultipliers (SiPMs) and is optimized to observe in the  $300 \lesssim \lambda/{\rm nm} \lesssim 900$ wavelength band for Cherenkov emission of showers developing towards the observatory. The PCC covers 9$^\circ$ of the FoV from the edge  (see Fig.~\ref{figComp}). The PCC SiPMs are assembled in arrays of 8 x 8 pixels with a total 15,360 pixels covering a 31 x 31 mm$^2$ area. With a time sampling of 20~ns, the PCC records EASs produced by cosmic rays above the limb of the Earth and showers from $\tau$-lepton decays below the Earth's limb induced by $\nu_\tau$ interactions in the Earth. 

The {\bf POEMMA mission} is designed for launch into a circular orbit at an inclination of 28.5$^\circ$ and an altitude of 525 km, implying a period of 95 mins. The satellites are launched in a stowed configuration. Once on orbit, the corrector plate and focal surface are deployed into their final position. After calibration, the instruments will be pointed close to the nadir to make stereo observations of UHECRs via fluorescent light. Once sufficient statistics have been acquired, the satellites separation will be reduced to $\sim$ 30 km and the instruments will be pointed for limb observations via both fluorescence and Cherenkov. Throughout the mission instruments will be re-oriented towards neutrino ToO directions following a transient event alert. During a ToO follow-up, measurements of fluorescence from EASs will continue utilizing the larger volumes of the atmosphere observed with the satellites pointed at the limb. 

\section{The Roadmap to POEMMA}

The design of the POEMMA observatory and mission evolved from previous work on the JEM-EUSO~\cite{JEM_EUSO_Overview} missions, the  OWL~\cite{OWL} design, and the CHANT~\cite{CHANT} concept. Among the JEM-EUSO missions,  Mini-EUSO and EUSO-SPB2 are paving the way to POEMMA. Mini-EUSO is measuring the low Earth orbit background for the fluorescence technique from the International Space Station~\cite{Mini-EUSO1,Mini-EUSO2}. The suborbital payload EUSO-SPB2~\cite{EUSO_SPB2_Overview} was designed to test the two main observation techniques of POEMMA with two separate telescopes: the EUSO-SPB2 Fluorescence Telescope (FT)~\cite{EUSO_SPB2_FT} and the Cherenkov Telescope (CT)~\cite{EUSO_SPB2_CT} . Both the FT and CT were designed as scaled down versions of the POEMMA optics for the 1 meter Schmidt optics of EUSO-SPB2. 

Figure~\ref{figComp} shows the correspondence between the two EUSO-SPB2 telescopes and POEMMA's PFC and PCC designs. The EUSO-SPB2 FT is a 1m diameter modified Schmidt telescope, with a 6,912 pixel camera of Multi-Anode Photomultiplier Tubes (MAPMTs) and integration time of 1$\mu$s. The MAPMTs are arranged in three photo-detector modules (PDMs) each composed of 36 MAPMTs  each with 8 × 8 pixels, resulting in a total of 2304 channels per PDM. (Mini-EUSO's camera consists of one PDM.) The FT field of view is 36$^\circ$ by 12$^\circ$ with a fixed nadir pointing direction.

The EUSO-SPB2 CT is also a 1m diameter modified Schmidt telescope with a bifocal alignment of the 4 mirror segments so signals are focused in two distinct spots separated by a couple of pixels on the camera to reduce the background of direct cosmic ray hits. The CT camera has a 512 pixel SiPM  camera with an integration time of 10~ns. The FoV of the instrument is 6.4$^\circ$ in zenith and 12.8$^\circ$ in azimuth and can be pointed during the flight from horizontal to 10$^\circ$ below the Earth's limb to follow ToOs.

EUSO-SPB2 was launched from Wanaka, New Zealand on May 13, 2023~\cite{EUSO_SPB2_Overview}. Although the flight was terminated early due to a leak in the super-pressure balloon, the data collected during the 36 hours and 53 mins flight show that the techniques of the POEMMA design can achieve the technical goals. The flight was too short for significant observations on EAS via fluorescence~\cite{EUSO_SPB2_FT}, but the CT was able to detect a number of events consistent with predicted high-energy cosmic rays~\cite{EUSO_SPB2_CT}. 

Two upcoming projects will further qualify POEMMA subsystems in orbital and sub-orbital platforms raising the technical readiness level of POEMMA: the Terzina~\cite{Terzina} Cherenkov detector on the NUSES small-satellite mission and the proposed POEMMA-Balloon-Radio (PBR) mission on a super-pressure balloon. Both are expected to launch in 2026.

\medskip
\noindent{\bf Acknowledgements}
The POEMMA conceptual study was supported by NASA Grant NNX17AJ82G. EUSO-SPB2 was supported by NASA awards 11-APRA-0058, 16-APROBES16-0023, 17-APRA17-0066, NNX13AH54G, 80NSSC18K0246, 80NSSC18K0473,80NSSC19K0626, 80NSSC18K0464, 80NSSC22K1488, 80NSSC19K0627 and 80NSSC22K0426, the French space agency CNES, the National Science Centre in Poland grant n. 2017/27/B/ST9/02162, and by ASI-INFN agreement n. 2021-8-HH.0 and its amendments. This research used resources of the US National Energy Research Scientific Computing Center (NERSC), the DOE Science User Facility operated under Contract No. DE-AC02-05CH11231.


\begin{thebibliography}{99}
\vspace{-0.3cm}
\bibitem{POEMMA_JCAP_2021} A. V. Olinto et al., JCAP 06 (2021) 007, arXiv:2012.07945
\vspace{-0.3cm}
\bibitem{UHECR_Snowmass} A. Coleman et al., Astropart. Phys. 149 (2023) 102819, arXiv:2205.05845
\vspace{-0.3cm}
\bibitem{UHECR_OpenQ} R. Alves Batista et al., Front.Astron.Space Sci. 6 (2019) 23, arXiv:1903.06714
\vspace{-0.3cm}
\bibitem{ANTARES:2017bia} 
  A.~Albert {\it et al.},  Astrophys.\ J.\  {\bf 850}, no. 2, L35 (2017), arXiv:1710.05839 
\vspace{-0.3cm}
\bibitem{Alvarez-Muniz:2018bhp}
J.~\'Alvarez-Mu\~niz \textit{et al.}, Sci. China Phys. M.\&A. \textbf{63} (2020) 1, 219501, arXiv:1810.09994 
\vspace{-0.3cm}
\bibitem{Fang:2017tla} 
  K.~Fang and B.~D.~Metzger, Astrophys.\ J.\  {\bf 849}, no. 2, 153 (2017), arXiv:1707.04263
\vspace{-0.3cm}
\bibitem{JEM_EUSO_Overview} M. Casolino, et al., in the Proceedings of the 38th International Cosmic Ray Conference, PoS (ICRC2023) 208 (2023)
\vspace{-0.3cm}
\bibitem{OWL} F. W. Stecker, et al., Nucl. Phys. Proc. Suppl. 136C, 433 (2004), astro-ph/0408162
\vspace{-0.3cm}
\bibitem{CHANT} A. Neronov, D. V. Semikoz, L. A. Anchordoqui, J. Adams and A. V. Olinto, Phys. Rev. D 95, no. 2,
023004 (2017), arXiv:1606.03629 
\vspace{-0.3cm}
\bibitem{Mini-EUSO1} L. Marcelli, et al., in the Proceedings of the 38th International Cosmic Ray Conference, PoS (ICRC2023) 001 (2023)
\vspace{-0.3cm}
\bibitem{Mini-EUSO2}
M.E. Bertaina, et al., in the Proceedings of the 38th International Cosmic Ray Conference, PoS (ICRC2023) 272 (2023) 
\vspace{-0.3cm}
\bibitem{EUSO_SPB2_Overview} J.B. Eser, et al., in the Proceedings of the 38th International Cosmic Ray Conference, PoS (ICRC2023) 397 (2023)
\vspace{-0.3cm}
\bibitem{EUSO_SPB2_FT}
G. Filippatos, et al., in the Proceedings of the 38th International Cosmic Ray Conference, PoS (ICRC2023) 251 (2023)
\vspace{-0.3cm}
\bibitem{EUSO_SPB2_CT}
E. Gazda et al., in the Proceedings of the 38th International Cosmic Ray Conference, PoS (ICRC2023) 1029 (2023)
\vspace{-0.3cm}
\bibitem{Terzina} R. Aloisio, et al., in the Proceedings of the 38th International Cosmic Ray Conference, PoS (ICRC2023) 391 (2023)
\vspace{-0.3cm}  


\end{thebibliography}

\newpage

{\Large\bf Full Authors list: The POEMMA Collaboration\\}

\begin{sloppypar}
{\small \noindent
J.H.~Adams Jr.,$^1$ 
R.~Aloisio,$^2$ 
L.A.~Anchordoqui,$^3$ 
A.~Anzalone,$^{4,5}$
M.~Bagheri,$^6$
D.~Barghini,$^7$
M.~Battisti,$^7$
D.R.~Bergman,$^8$ 
M.E.~Bertaina,$^7$ 
P.F.~Bertone,$^9$ 
F.~Bisconti,$^{10}$
M.~Bustamante,$^{11}$
F.~Cafagna,$^{12}$
R.~Caruso,$^{13,5}$
M.~Casolino,$^{14,15,16}$ 
K.~\v{C}ern\'{y},$^{17}$
M.J.~Christl,$^{9}$
A.L.~Cummings,$^{18}$  
I.~De Mitri,$^2$
R.~Diesing,$^{19}$
R.~Engel,$^{20}$
J.B.~Eser,$^{19}$ 
K.~Fang,$^{21}$
F.~Fenu,$^{22}$ 
G.~Filippatos,$^{23}$ 
E.~Gazda,$^6$
C.~Gu\'epin,$^{24}$ 
A.~Haungs,$^{20}$
E.A.~Hays,$^{25}$  
E.G.~Judd,$^{26}$   
P.~Klimov,$^{27}$ 
J.F.~Krizmanic,$^{25}$ 
V.~Kungel,$^{23}$ 
E.~Kuznetsov,$^1$
\v{S}.~Mackovjak,$^{28}$  
D.~Mand\'{a}t,$^{29}$ 
L.~Marcelli,$^{14}$
J.~McEnery,$^{25}$ 
G.~Medina-Tanco,$^{30}$ 
K.-D.~Merenda,$^{23}$ 
S.~S.~Meyer,$^{19}$
J.W.~Mitchell,$^{25}$ 
H.~Miyamoto,$^7$ 
J.~M.~Nachtman,$^{31}$
 A.~Neronov,$^{32,33}$ 
 F.~Oikonomou,$^{34}$
 A.V.~Olinto, $^{19}$
 Y.~Onel,$^{31}$
G.~Osteria,$^{35}$
A.N.~Otte,$^{6}$ 
E.~Parizot,$^{33}$ 
T.C.~Paul,$^{3}$ 
M.~Pech,$^{29}$
J.S.~Perkins,$^{25}$ 
P.~Picozza,$^{14,15}$ 
L.W.~Piotrowski,$^{36}$ 
Z.~Plebaniak,$^7$
G.~Pr\'ev\^ot,$^{33}$ 
P. Reardon,$^{1}$ 
M.H.~Reno,$^{31}$
M.~Ricci,$^{37}$
O.~Romero Matamala,$^{6}$ 
F.~Sarazin,$^{23}$ 
P.~Schov\'{a}nek,$^{29}$
V.~Scotti,$^{35,38}$
K.~Shinozaki,$^{39}$ 
J.F.~Soriano,$^3$
F.~Stecker,$^{25}$ 
Y.~Takizawa,$^{16}$
M.~Unger,$^{20}$
T.M.~Venters,$^{25}$  
L.~Wiencke,$^{23}$ 
D.~Winn,$^{31}$
R.M.~Young,$^{9}$
M.~Zotov.$^{27}$ 
}
\end{sloppypar}

{\footnotesize
\noindent
$^1$University of Alabama in Huntsville, Huntsville, AL, USA;\\
$^2$Gran Sasso Science Institute, L'Aquila, Italy;\\
$^3$City University of New York, Lehman College, NY, USA;\\
$^4$Istituto Nazionale di Astrofisica INAF-IASF, Palermo, Italy;\\
$^5$Istituto Nazionale di Fisica Nucleare, Catania, Italy;\\
$^6$Georgia Institute of Technology, Atlanta, GA, USA;\\
$^7$Universita' di Torino, Torino, Italy;\\
$^8$University of Utah, Salt Lake City, Utah, USA;\\
$^9$NASA Marshall Space Flight Center, Huntsville, AL, USA;\\
$^{10}$Istituto Nazionale di Fisica Nucleare, Sezione di Torino, Italy;\\
$^{11}$Niels Bohr Institute, University of Copenhagen, Copenhagen, Denmark;\\
$^{12}$Istituto Nazionale di Fisica Nucleare, Bari, Italy;\\
$^{13}$Universita' di Catania, Catania, Italy;\\
$^{14}$Istituto Nazionale di Fisica Nucleare, Sezione di Roma Tor Vergata, Italy;\\
$^{15}$Universit\`a di Roma Tor Vergata, Roma, Italy;\\
$^{16}$RIKEN, Wako, Japan;\\
$^{17}$Joint Laboratory of Optics, Faculty of Science, Palack\'{y} University, Olomouc, Czech Republic;\\
$^{18}$Pennsylvania State University, PA, USA;\\
$^{19}$The University of Chicago, Chicago, IL, USA;\\
$^{20}$Karlsruhe Institute of Technology, Karlsruhe, Germany;\\
$^{21}$University of Wisconsin, Madison, WI, USA;\\
$^{22}$Agenzia Spaziale Italiana, Via del Politecnico, 00133, Roma, Italy;\\
$^{23}$Colorado School of Mines, Golden, CO, USA;\\
$^{24}$Laboratoire Univers et Particules de Montpellier, Montpellier, France;\\
$^{25}$NASA Goddard Space Flight Center, Greenbelt, MD, USA;\\
$^{26}$Space Sciences Laboratory, University of California, Berkeley, CA, USA;\\
$^{27}$Skobeltsyn Institute of Nuclear Physics, Lomonosov Moscow State University, Moscow, Russia;\\
$^{28}$Institute of Experimental Physics, Slovak Academy of Sciences, Kosice, Slovakia;\\
$^{29}$Institute of Physics of the Czech Academy of Sciences, Prague, Czech Republic;\\
$^{30}$Instituto de Ciencias Nucleares, UNAM, CDMX, Mexico;\\
$^{31}$University of Iowa, Iowa City, IA, USA;\\
$^{32}$University of Geneva, Geneva, Switzerland;\\
$^{33}$Universit\'e de Paris, CNRS, Astroparticule et Cosmologie, F-75013 Paris, France;\\
$^{34}$Norwegian University of Science and Technology, Trondheim, Norway;\\
$^{35}$Istituto Nazionale di Fisica Nucleare, Napoli, Italy;\\
$^{36}$Faculty of Physics, University of Warsaw, Warsaw, Poland;\\
$^{37}$Istituto Nazionale di Fisica Nucleare - Laboratori Nazionali di Frascati, Frascati, Italy;\\
$^{38}$Universita' di Napoli Federico II, Napoli, Italy;\\
$^{39}$National Centre for Nuclear Research, Otwock, Poland.\\
}
\bigskip 

{\Large\bf Full Authors list: The JEM-EUSO Collaboration\\}

\begin{sloppypar}
{\small \noindent
S.~Abe$^{ff}$, 
J.H.~Adams Jr.$^{ld}$, 
D.~Allard$^{cb}$,
P.~Alldredge$^{ld}$,
R.~Aloisio$^{ep}$,
L.~Anchordoqui$^{le}$,
A.~Anzalone$^{ed,eh}$, 
E.~Arnone$^{ek,el}$,
M.~Bagheri$^{lh}$,
B.~Baret$^{cb}$
% ...
}
\end{sloppypar}

{ \footnotesize
\noindent
% ...
$^{cb}$ Universit\'e de Paris, CNRS, AstroParticule et Cosmologie, F-75013 Paris, France\\
% ...
$^{ed}$ Istituto Nazionale di Fisica Nucleare - Sezione di Catania, Italy\\
% ...
$^{eh}$ INAF - Istituto di Astrofisica Spaziale e Fisica Cosmica di Palermo, Italy\\
% ...
$^{ek}$ Istituto Nazionale di Fisica Nucleare - Sezione di Torino, Italy\\
$^{el}$ Dipartimento di Fisica, Universit\`a di Torino, Italy\\
% ...
$^{ep}$ Gran Sasso Science Institute, L'Aquila, Italy\\
% ...
$^{ff}$ Nihon University Chiyoda, Tokyo, Japan\\ 
% ...
$^{ld}$ University of Alabama in Huntsville, Huntsville, AL, USA\\
$^{le}$ Lehman College, City University of New York (CUNY), NY, USA\\
% ...
$^{lh}$ Georgia Institute of Technology, USA\\
% ...
}

%

\end{document}